\documentclass[prl, twocolumn, superscriptaddress, preprintnumbers, showpacs, longbibliography ]{revtex4-1}
\pdfoutput=1
\usepackage[T1]{fontenc}
\usepackage{bm}
\usepackage{subfigure}
\usepackage{amssymb}
\usepackage{graphicx}
\usepackage{amsmath}
\usepackage{tensor}
\usepackage{epstopdf}
\usepackage{extarrows}
\usepackage{xcolor}
\usepackage[colorlinks=true,
              linkcolor=blue,
              urlcolor=blue,
              citecolor=blue
              ]{hyperref}

\newcommand{\be}{\begin{equation}}
\newcommand{\ee}{\end{equation}}
\newcommand{\bea}{\setlength\arraycolsep{2pt} \begin{eqnarray}}
\newcommand{\eea}{\end{eqnarray}}
\newcommand{\nn}{\nonumber}
\newcommand{\mm}{\mathrm}
\newcommand{\mc}{\mathcal}

\def\ft#1#2{{\textstyle{\frac{\scriptstyle #1}{\scriptstyle #2} } }}
\def\fft#1#2{{\frac{#1}{#2}}}

\def\0{{\sst{(0)}}}
\def\1{{\sst{(1)}}}
\def\2{{\sst{(2)}}}
\def\3{{\sst{(3)}}}
\def\4{{\sst{(4)}}}
\def\5{{\sst{(5)}}}
\def\6{{\sst{(6)}}}
\def\7{{\sst{(7)}}}
\def\8{{\sst{(8)}}}
\def\sst#1{{\scriptscriptstyle #1}}

\begin{document}

\title{Topological interpretation for phase transitions of black holes}

\author{Zhong-Ying Fan}
\email{fanzhy@gzhu.edu.cn}
\affiliation{Department of Astrophysics, School of Physics and Materials Science, Guangzhou University, Guangzhou 510006, China }

\begin{abstract}
In this work, we develop a universal picture from topology in the thermodynamic parameters space to describe first order phase transitions of black holes. By employing an off-shell internal energy, we find two types of topological defects. The first is normal, describing black holes which move with a nonzero velocity and acceleration, in a central force field. The second type of defects are exotic: they are static in the space, not describing black holes but encoding information about first order transitions. For each defect, we assign a winding number and an inertial mass. By studying neutral and charged black holes in asymptotically anti-de Sitter space, we show that first order transitions can be viewed as once or twice interchange of winding numbers between black holes and the exotic defects, through wired action at a distance. This corresponds to the usual notion: a smaller black hole grows into a larger black hole or vice versa. However, our topological analysis illustrates that the transition can also be locally interpreted as virtual collisions between black holes and the exotic defects. In this interpretation, a smaller black hole first grows into a new exotic defect whereas an original exotic defect grows into a larger black hole. All the defects simply change their positions and momentums, rather than interchanging the winding numbers. Critical point of the transition can be extracted when all the defects meet in the parameters space. Certain quantities, such as the Jacobians and the velocities of normal defects show universal behaviors near the critical point.



\end{abstract}
\maketitle

\emph{Introduction.---} Topology is an efficient tool to study generic properties of a physical system. It has been widely applied to particle physics and condensed matter systems since a long time ago. In 1980s, Duan developed a phenomenological theory to systematically study topological current and charge for a physical system \cite{Duan:1984ws}. The theory predicted very interesting results about kinetics of topological defects \cite{Duan:1998kw,Fu:2000pb}, such as collisions, coalesces and splits of defects and generation/annihilation of a pair of vortex-antivortex.

Recently, the theory has been adopted to study certain properties of gravitational objects, such as light rings \cite{Cunha:2017qtt,Cunha:2020azh,Guo:2020qwk,Wei:2020rbh} and thermodynamic properties of black holes \cite{Wei:2021vdx,Yerra:2022alz,Yerra:2022eov,Ahmed:2022kyv,Yerra:2022coh,Wei:2022dzw}. However, in all these works, phase transitions of black holes were not studied carefully from topology.


In fact, first order phase transitions are very common phenomena in nature. Some of the features are believed to be universal, for example critical behaviors of physical quantities near the critical point. Inspired by this, in this work, we employ an off-shell internal energy to develop a universal picture from topology in the thermodynamic parameters space to describe first order transitions for macroscopic systems in equilibrium. We mainly focus on black holes since these objects are particularly simple, characterized by only a few quantities. A particularly interesting example is charged black holes in asymptotically anti-de Sitter (AdS) space. By taking cosmological constant as the thermodynamic pressure, it was shown \cite{Kubiznak:2012wp} that the small-large black hole transition is very similar to the liquid-gas transition of Van der Waals fluids. This provides us a good example to test our prescription.

\emph{Phenomenological theory.---}A key concept of topology is that of defects, which are defined as the zero points of a vector field in a space
\be \phi^a(\tau\,,\vec{x})\Big|_{\vec{x}=\vec{z}}=0 \,,\quad \vec{z}=\vec{z}(\tau)\,,\label{defect}\ee
where $\vec{z}$ denotes the position of a defect, which generally moves in the space. The simplest topological quantity associated to a defect is its winding number.  To be specific,  we focus on a two-component vector field: $\phi^a=(\phi^1\,,\phi^2)$, defined in the coordinate space $x^\mu=(\tau\,,x^1\,,x^2)$. To describe phase transitions of black holes, we take $x^1=r_h$, being the horizon radii and $x^2=\theta$, which is a parameter introduced by hand for convenience. The meaning of the time coordinate $\tau$ will be clarified later when it is solved from the condition of defects.

We introduce an off-shell internal energy
\be \mc{E}=F+\tau/\beta\,,\label{offenergy}\ee
where $F$ is the free energy evaluated from the Euclidean gravitational action and $\beta=T^{-1}$ is the inverse of temperature. The vector field $\phi^a$ is defined as
\be\label{phi} \phi^{1}=-\fft{\partial\mc{E}}{\partial r_h}\,,\quad \phi^{2}=-\mm{cot}\theta\, \mm{csc}\theta \,,\ee
where $0<\tau<+\infty\,,0<r_h<+\infty $ and $ 0\leq \theta<\pi$.

According to Duan's $\phi$-mapping theory, a topological current $j^\mu$ can be defined as \cite{Duan:1998kw,Fu:2000pb}
\be\label{tpcurrent} j^\mu=\fft{1}{2\pi}\epsilon^{\mu\nu\rho}\epsilon_{ab}\partial_\nu n^a \partial_\rho n^b \,,\ee
where $n^a=\phi^a/\phi$ and $\phi$ is the norm of $\phi^a$. It is obvious that the current $j^\mu$ is identically conserved: $\partial_\mu j^\mu=0$.
It turns out that the topological current $j^\mu$ is of a delta function of the field configuration \cite{Duan:1998kw,Fu:2000pb}
\be j^\mu=\delta^2(\vec{\phi})J^\mu(\ft{\phi}{x}) \,,\label{current}\ee
where $J^\mu(\ft{\phi}{x})$ is a three dimensional Jacobian defined as
$ \epsilon^{ab}J^\mu({\ft{\phi}{x}})=\epsilon^{\mu\nu\rho}\partial_\nu\phi^a\partial_\rho\phi^b$.
Using ordinary theory of $\delta$-function and
considering the case that while the point $\vec{x}$ covers the region neighboring the zero $\vec{x}=\vec{z}_\alpha$ once, the $\phi$-field covers the corresponding region $\beta_\alpha$ times, one has
\be j^0=\sum_\alpha \beta_\alpha \eta_\alpha\, \delta^2(\vec{x}-\vec{z}_\alpha) \,,\label{j0}\ee
where $\beta_\alpha$ is the Hopf index and $\eta_\alpha=\big[ J^0(\ft{\phi}{x})/|J^0(\ft{\phi}{x})|\big]_{\vec{x}=\vec{z}_\alpha}=\pm 1 $,
is the Brouwer degree. The winding number associated to the zero $\vec{x}=\vec{z}_\alpha$ is $w_\alpha=\beta_\alpha\eta_\alpha$. The global topological charge in a region $V$ is given by
\be W=\int_V \sum_\alpha w_\alpha \,\delta^2(\vec{x}-\vec{z}_\alpha)\,d^2x=\sum_\alpha w_\alpha  \,.\label{inner}\ee
This characterizes the inner structure of defects in the region $V$.

The velocity of the defect at $\vec{x}=\vec{z}_\alpha$ can be evaluated as $u^i=\big[J^i(\ft{\phi}{x})/J^0(\fft{\phi}{x})\big]_{\vec{x}=\vec{z}_\alpha}$. However, when the Jacobian $J^0(\ft{\phi}{x})=0$ at a certain zero $(\tau_*\,,z_*^1)$, a defect will bifurcate. This usually happens when the defects have multiple solutions. Duan's $\phi$-mapping theory predicted rich results about kinetics of defects in this case \cite{Duan:1998kw,Fu:2000pb}. The relevant case to us is the Jacobian $J^1(\ft{\phi}{x})=0$ as well at $(\tau_*\,,z_*^1)$. In this case, the functional relation between $x^1$ and $\tau$ will not be unique in the neighborhood of the bifurcation point. Instead, the defects are determined by Taylor expansion of $\phi^1(\tau\,,x^1)$ around $(\tau_*\,,z_*^1)$
\be A\Big(\fft{dx^1}{d\tau}\Big)^2+2B\fft{dx^1}{d\tau}+C=0 \,,\label{bif}\ee
where
\be A=\fft{\partial^2\phi^1}{(\partial x^1)^2}\Big|_{(\tau^*\,,z_*^1)}\,, B=\fft{\partial^2\phi^1}{\partial x^1\partial \tau}\Big|_{(\tau^*\,,z_*^1)}\,, C=\fft{\partial^2\phi^1}{\partial \tau^2}\Big|_{(\tau^*\,,z_*^1)} \,.\nn\ee
Generally the above equation has two branch solutions, characterizing collisions, coalesces or splits of defects at the bifurcation point. However, if the point has a higher degeneracy, there will exist more branch solutions, determined by expansion of $\phi^1(t\,,x^1)$ to higher orders.

\emph{Topological defects and phase transition.---}By definition (\ref{phi}), the zero points of $\phi$-vector field require $\theta=\pi/2$ and generally the time coordinate $\tau$ gives the Bekenstein-Hawking entropy $\tau=-\partial F/\partial T=S$. This describes black holes. However, there exist un-black hole solutions. We deduce
\be \phi^1=(\pi r_h^2-\tau)J^1\,,\quad J^1=\partial T/\partial r_h\,.\label{phi1}\ee
 Clearly, $J^1=0$ gives alternate solutions of defects, leaving the time coordinate $\tau$ undetermined. This new branch solutions describe static defects, referred to as exotic ones. Existence of this type of defects is crucial to describe first order transitions.

To proceed, we evaluate the three dimensional Jacobians at the defects (not necessarily for black holes)
\bea
J^1&=&\partial_{r_h}T=2\pi T r_h /C_T \,,\nn\\
J^0&=&2\pi r_h J^1+(\pi r_h^2-\tau)\partial_{r_h}J^1\,,
\label{jacob}\eea
where $C_T=T dS/dT$ is the specific heat (defined at constant pressure). Note $J^2$ identically vanishes and the results in this direction are trivial. For black holes, $\tau=\pi r_h^2$, one finds $J^0=4\pi T S/C_T$. Remarkably, the Jacobian $J^0$ shares the same sign with the specific heat as well as the winding number according to (\ref{j0}). It implies that a positive/negative winding number means the black hole is locally (meta)stable/unstable in thermodynamics. Since the specific heat $C_T$ diverges at the critical point of first order transitions, one finds $J^1=0=J^0$ for black holes and $J^1=0\,,J^0=(\pi r_{h*}^2-\tau_*)\partial_{r_{h*}}J^1$ for the exotic defects. The two branch solutions will catch up only at the bifurcation points at which $\tau_*=\pi r_{h*}^2$.
This tells us that the bifurcation points play important roles in first order transitions.

According to (\ref{jacob}), a normal defect generally moves with a velocity $u^1=1/2\pi r_h\neq 0$. The result is universal to black holes. The acceleration is given by $a^1=-1/4\pi^2 r^3_h$, which implies the normal defect moves as a point particle in a central force field. This inspires us to introduce an inertial mass $m$ for normal defects to characterize the energy transfer during first order transitions. A natural requirement is the latent heat $L$ equals to the increase of potential energy associated to the central force field: $ L=\Delta V$, where $V=-m/8\pi^2r_h^2$.
Note the total mechanical energy is always zero. In canonical ensemble, the inertial mass can be evaluated as
\be m(T)=8\pi^3 T r^2_{h\,,s}r^2_{h\,,l} \,.\label{imass}\ee
The mass is simply a function of temperature, implying that a first order transition occurs between a pair of black holes which have the same inertial mass. For exotic defects, we assign a zero mass since they are not attracted by the central force field.

\emph{Examples--}
 Let us study several examples. The first is neutral black holes in AdS space. The equation of state is given by
\be T=\fft{1+3r_h^2\ell^{-2} }{4\pi r_h}\,,\ee
where $\ell$ is AdS radius. It is well known that in AdS spacetime there exists a minimal temperature $T_c=\sqrt{3}/2\pi\ell$ above which thermal radiation is unstable against gravitational collapse to form a black hole, i.e. the Hawking-Page transition \cite{Hawking:1982dh}. Above $T_c$, there exists a pair of black holes for a given temperature. The smaller/larger black hole is unstable/stable because of a negative/positive specific heat. There is a first order transition between the two black holes but the two phases do not coexist.
\begin{figure}\label{sch}
  \centering
  \includegraphics[width=180pt]{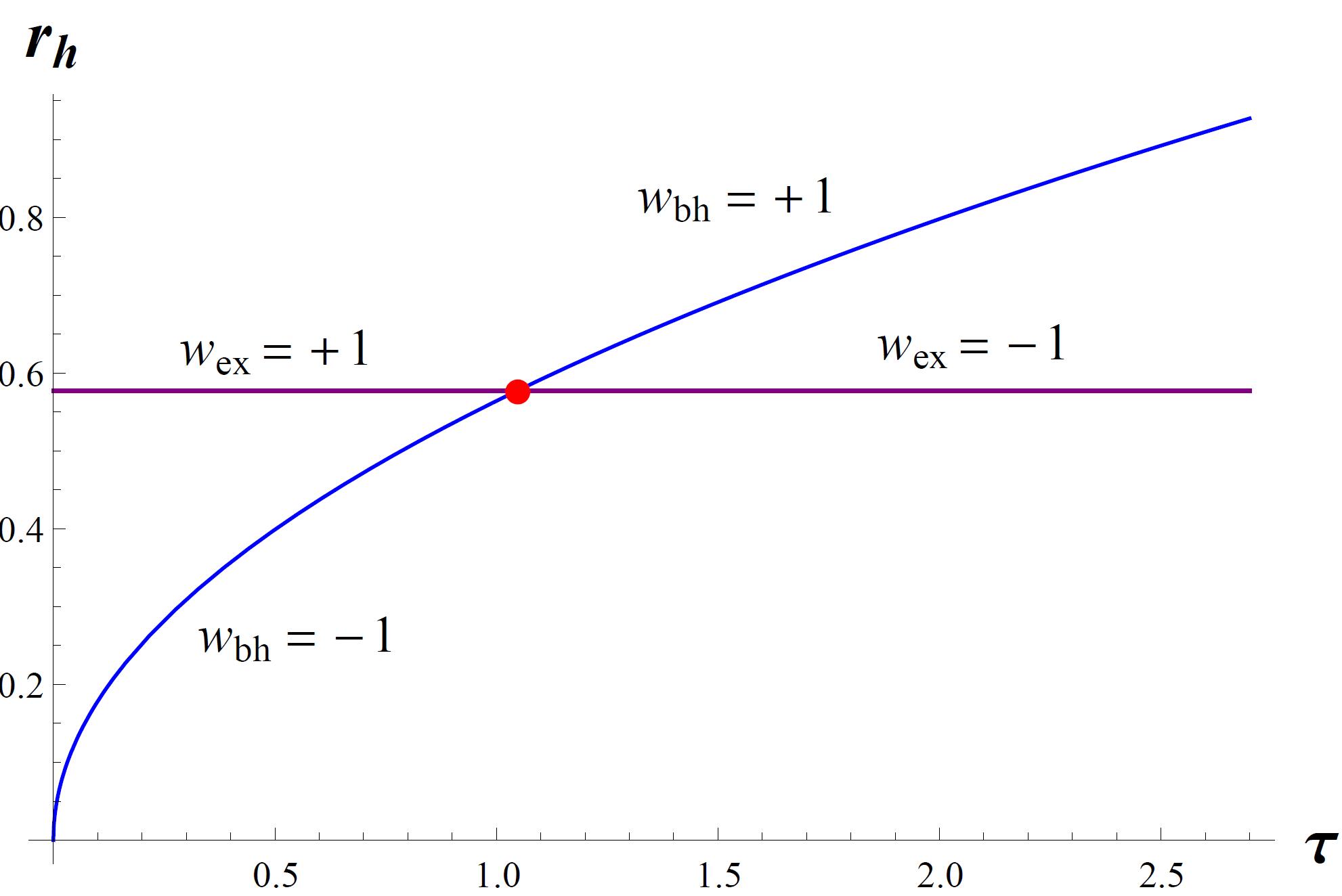}
  \caption{Defects for neutral black holes in AdS. The blue/horizontal line stands for black holes/exotic defects. The bifurcation point $\tau_*=\pi r_{h*}^2$ is represented by the red point, at which a smaller black hole collides with the exotic defect.}\label{sch}
\end{figure}

The vector field component $\phi^1$ is given by (\ref{phi1}), with Jacobians
\bea\label{schjacob}
J^1=\fft{3r_h^2\ell^{-2}-1}{4\pi r_h^2}\,,\quad
J^0=\fft{3\pi r_h^4\ell^{-2}-\tau}{2\pi^2 r_h^3}\,.
\eea
 Clearly there are two types of defects. The first describes black holes $\tau=\pi r_h^2$, with a velocity $u^1=1/2\pi r_h$. The second corresponds to $J^1=0$, leading to $r_h=\ell/\sqrt{3}$. This describes a static defect since $u^1=0$. However, there exists a bifurcation point $(\tau_*\,,r_{h*})$, at which $J^0=0=J^1$. One finds $r_{h*}=\ell/\sqrt{3}$ and $\tau_*=\pi r_{h*}^2$. Substituting the result into (\ref{bif}), we find the velocity of defects at the bifurcation point is given by $u^1_*=0$ or $u^1_*=1/2\pi r_{h*}$.
This implies that the black hole and the exotic defect collide at the bifurcation point and then depart, see Fig.\ref{sch}. Apparently, after the collision, the defects do not change their velocities but interchange their local winding numbers. According to (\ref{schjacob}), when $\tau<\tau_*$, a smaller black hole has $w_{bh}=-1$ whereas the exotic defect has $w_{ex}=+1$. Of course, the black hole is unstable. However, when $\tau>\tau_*$, a larger black hole has $w_{bh}=+1$, becoming stable but now $w_{ex}=-1$. In any case, the global topological charge remains zero owing to the conservation law. In the usual notion, a smaller black hole grows into a larger black hole. This implies the smaller black hole interchanges its winding number with the exotic defect during the transition. Critical point of the transition exactly corresponds to the bifurcation point, at which the defects meet in the parameters space.

\begin{figure}
  \centering
  \includegraphics[width=120pt]{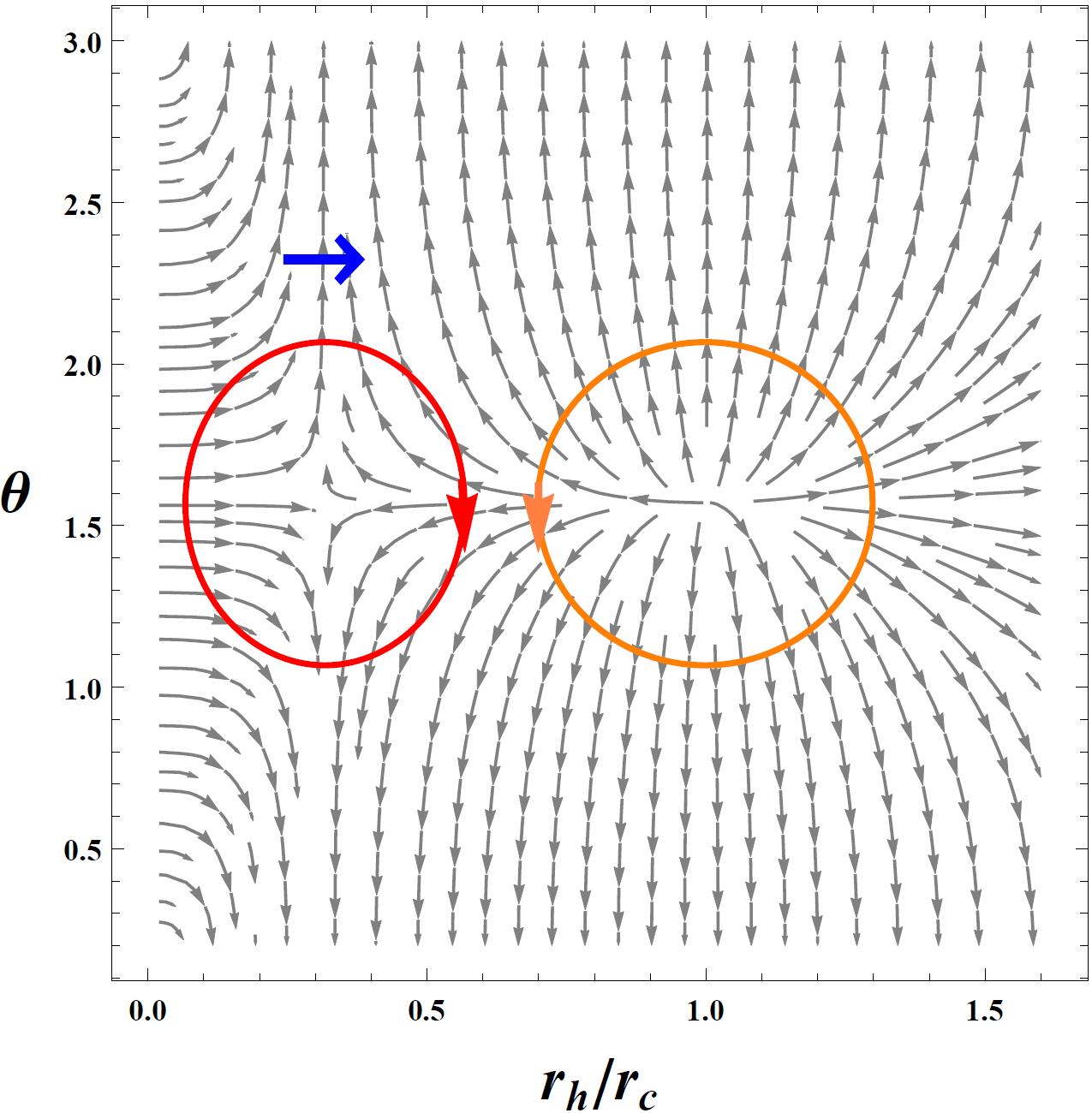}
  \includegraphics[width=120pt]{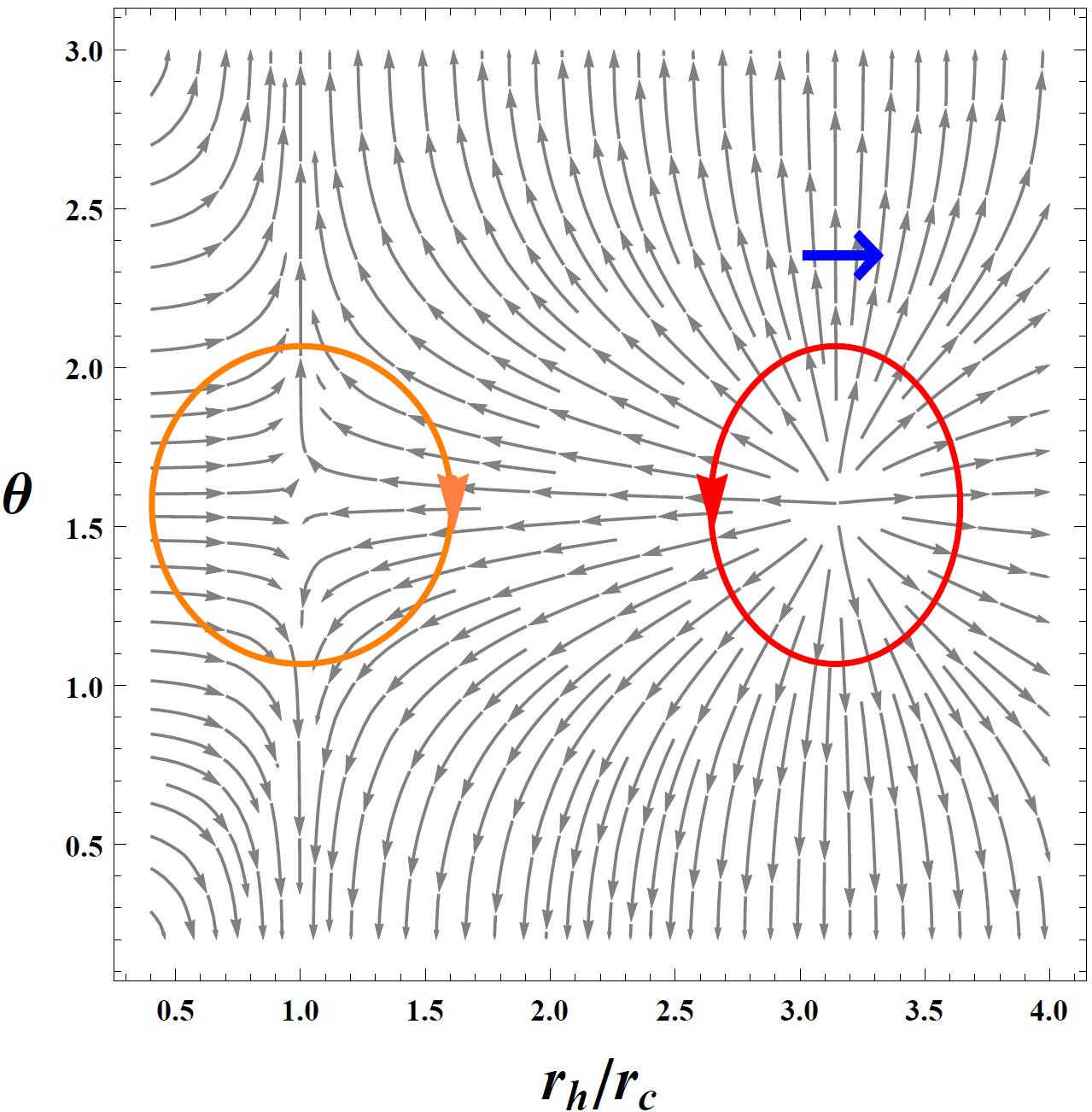}
  \caption{Collisions between defects for neutral black holes at temperature $T=\sqrt{3}T_c$. The stream plot is given for the unit vector field $n^a$. The red/orange circles stand for black holes/exotic defects. In the left panel, a smaller black hole collides with the exotic defect, interchanging their momentums. In the right panel, the smaller black hole becomes an exotic defect whereas the original one becomes a larger black hole. }
  \label{schdef}
\end{figure}

However, this interpretation relies on the action at a distance between the defects for temperature $T>T_c$. However, according to the $\phi$-field configuration shown in Fig.\ref{schdef}, a local explanation for the transition is the smaller black hole collides with the exotic defect, interchanging their momentums. Then the black hole itself becomes an exotic defect whereas the original one becomes a larger black hole. The process is similar to the elastic scattering of identical particles in classical mechanics. The net effect is the defects simply change their positions and momentums rather than interchanging their winding numbers. Since the temperature is fixed during the transition, the collision between the defects should be viewed as virtual.




Our second example is charged black holes in AdS spacetime. The equation of state is given by \cite{Kubiznak:2012wp}
\be P=\fft{T}{2r_h}-\fft{1}{8\pi r_h^2}+\fft{Q^2}{8\pi r_h^4}\,,\ee
where $P\equiv -\Lambda/8\pi$, is taken as the thermodynamic pressure. It was shown \cite{Kubiznak:2012wp} that below a critical temperature $T_c$, there exists a small-large black hole transition, along the coexistence curve \cite{Spallucci:2013osa}:
$ \tilde{T}^2=\tilde{P}\big(3-\sqrt{\tilde P} \big)/2$,
where $\tilde{T}=T/T_c\,,\tilde{P}=P/P_c$. Now in our definition (\ref{offenergy}), $F$ should be identified to the Gibbs free energy and all derivatives with respect to $r_h$ should be evaluated at constant $P\,,Q$.

\begin{figure}
  \centering
  \includegraphics[width=120pt]{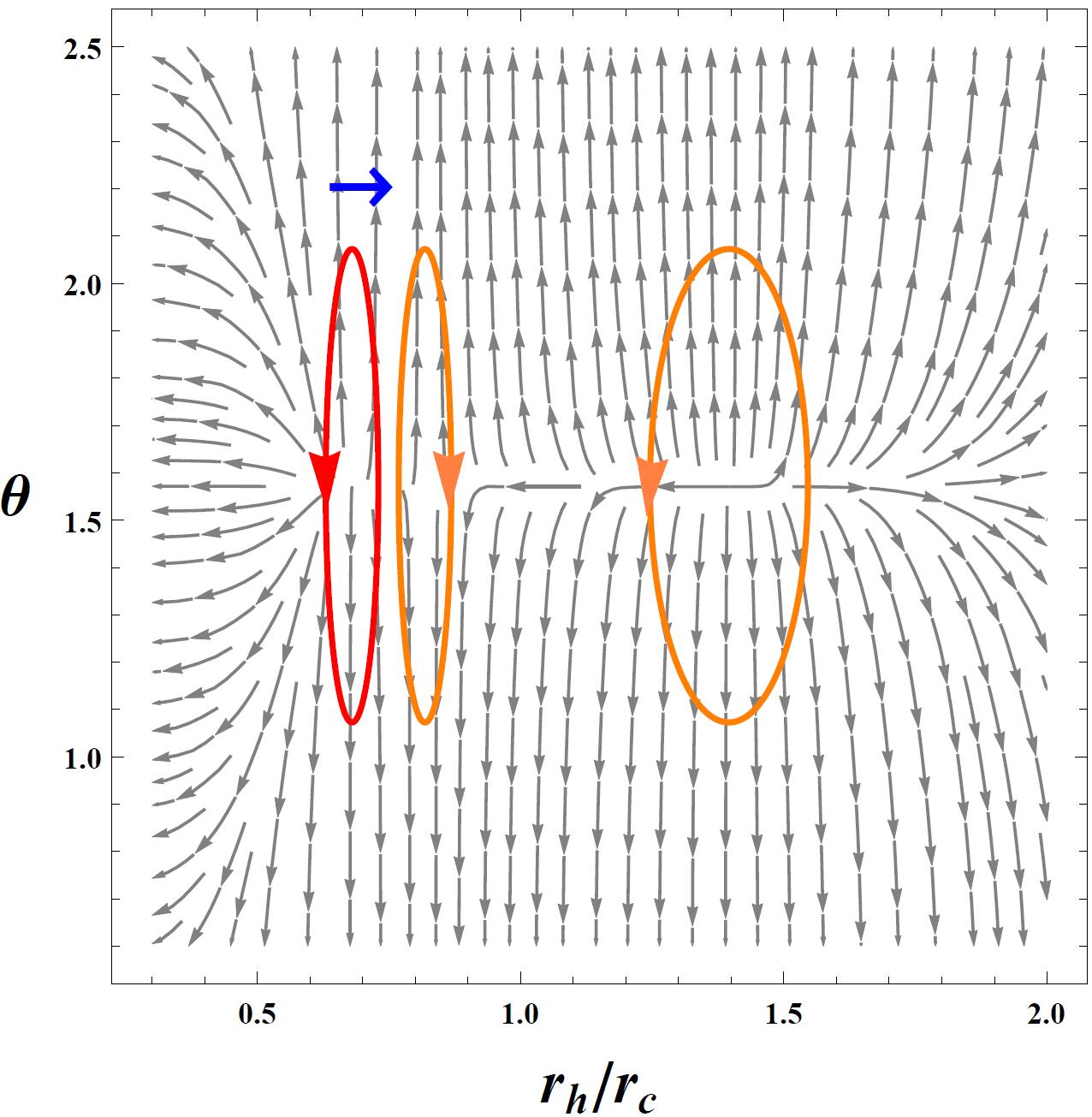}
  \includegraphics[width=120pt]{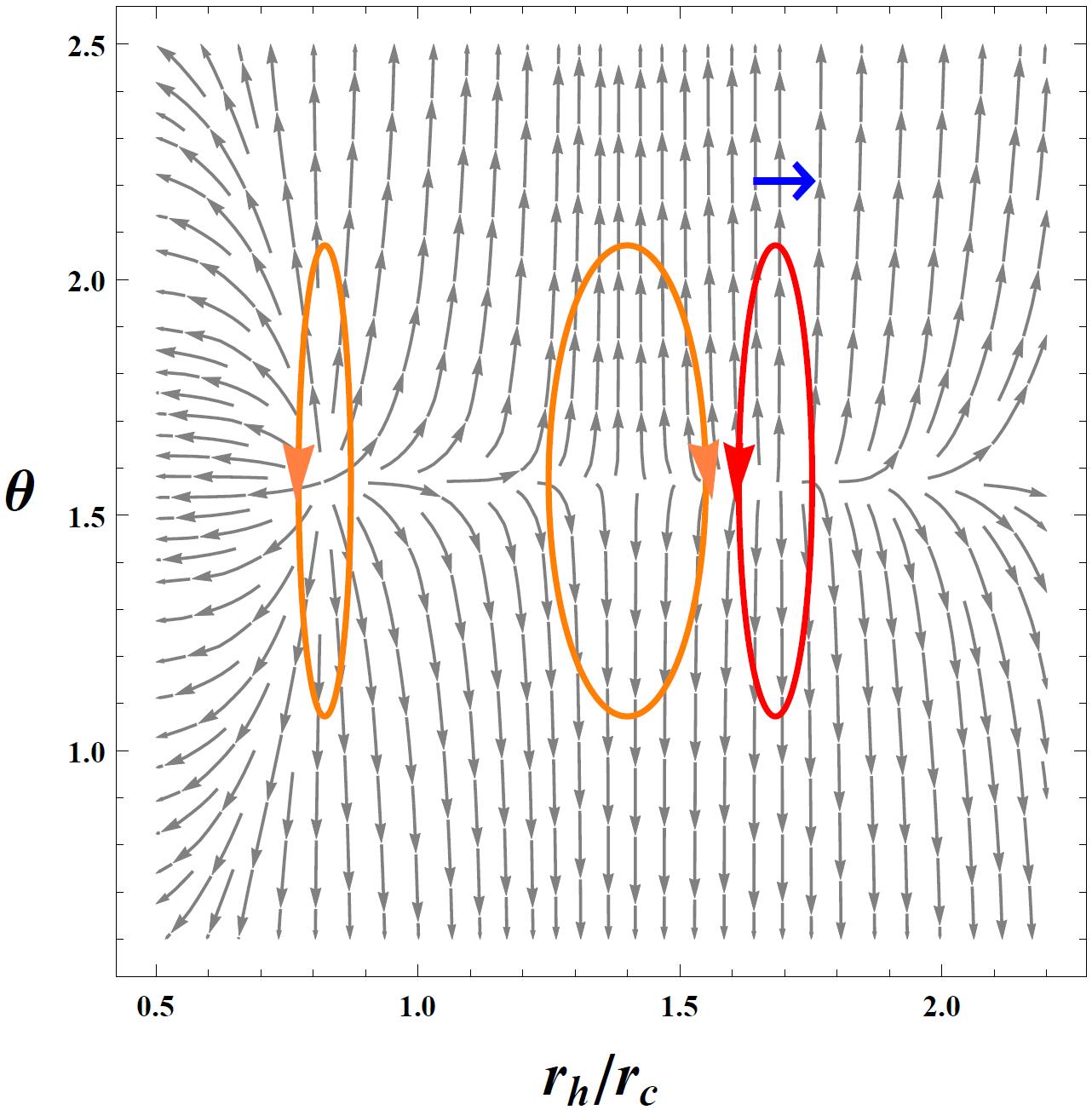}
  \caption{Collisions between defects for charged AdS black holes at temperature $T=9T_c/10$. The stream plot is given for the unit vector field $n^a$. The red/orange circles stand for black holes/exotic defects. In the left panel, a smaller black hole collides with a smaller exotic defect, interchanging their momentums. The latter continues colliding with the larger exotic defect elastically. In the right panel, the smaller black hole becomes a new smaller exotic defect while the original smaller one becomes a new larger exotic defect. The original larger one becomes a larger black hole. }
  \label{rndef}
\end{figure}

The Jacobians are given by
\bea\label{chargedjacob}
&&J^1=\fft{8\pi P\,r_h^4-r_h^2+3Q^2}{4\pi r_h^4}  \,,\\
&&J^0=\fft{8\pi^2 P\,r_h^6-3\pi Q^2 r_h^2-(r_h^2-6Q^2)\tau}{2\pi r_h^5}\,.\nn
\eea
In general, there are two types of defects. The first is black holes described by $\tau=\pi r_h^2$, with the velocity $u^1=1/2\pi r_h$. The second type is exotic, determined by $J^1=0$ and has a vanishing velocity $u^1=0$. However, existence of the exotic defects requires $P<1/96\pi Q^2$. Otherwise, no such defects exist. From (\ref{chargedjacob}), we obtain for the exotic defects
\be r_{h\pm}^2=\fft{1\pm \sqrt{1-96\pi  P Q^2}}{16\pi P} \,.\ee
As the pressure $P$ varies, we obtain one parameter family of a pair of exotic defects until $P=1/96\pi Q^2$. The bifurcation points are located at $r_h^*=r_{h\pm}$ and $\tau^*=\pi r_{h}^{*2}$, where $J^1=0=J^0$. As the previous case, at each bifurcation point, a black hole collides with an exotic defect and then departs, without changing their velocities, as shown in Fig.\ref{rn}. When $T<T_c$, this illustrates a twice topological charge interchange picture for the small-large black hole transition. It corresponds to the isobar in the $P-r_h$ diagram, meaning a smaller black hole grows into a larger black hole or vice versa. However, according to the $\phi$-field configuration shown in Fig.\ref{rndef}, the transition can also be locally interpreted as: the smaller black hole first collides with the smaller exotic defect, interchanging their momentums. Then the black hole itself reduces to a new exotic defect whereas the smaller exotic defect continues colliding with the larger exotic defect elastically. After the collision, the smaller exotic defect becomes a new larger exotic defect whereas the original larger one becomes a larger black hole. In this process, all the defects simply change their positions and momentums rather than interchanging the winding numbers. Since the specific volume of black holes jumps discontinuously during the transition, the process should be viewed as virtual. This interpretation corresponds to the oscillatory region in the $P-r_h$ diagram. Equivalence between the two explanations is guaranteed by the Maxwell's area law.


\begin{figure}\label{rn}
  \centering
  \includegraphics[width=180pt]{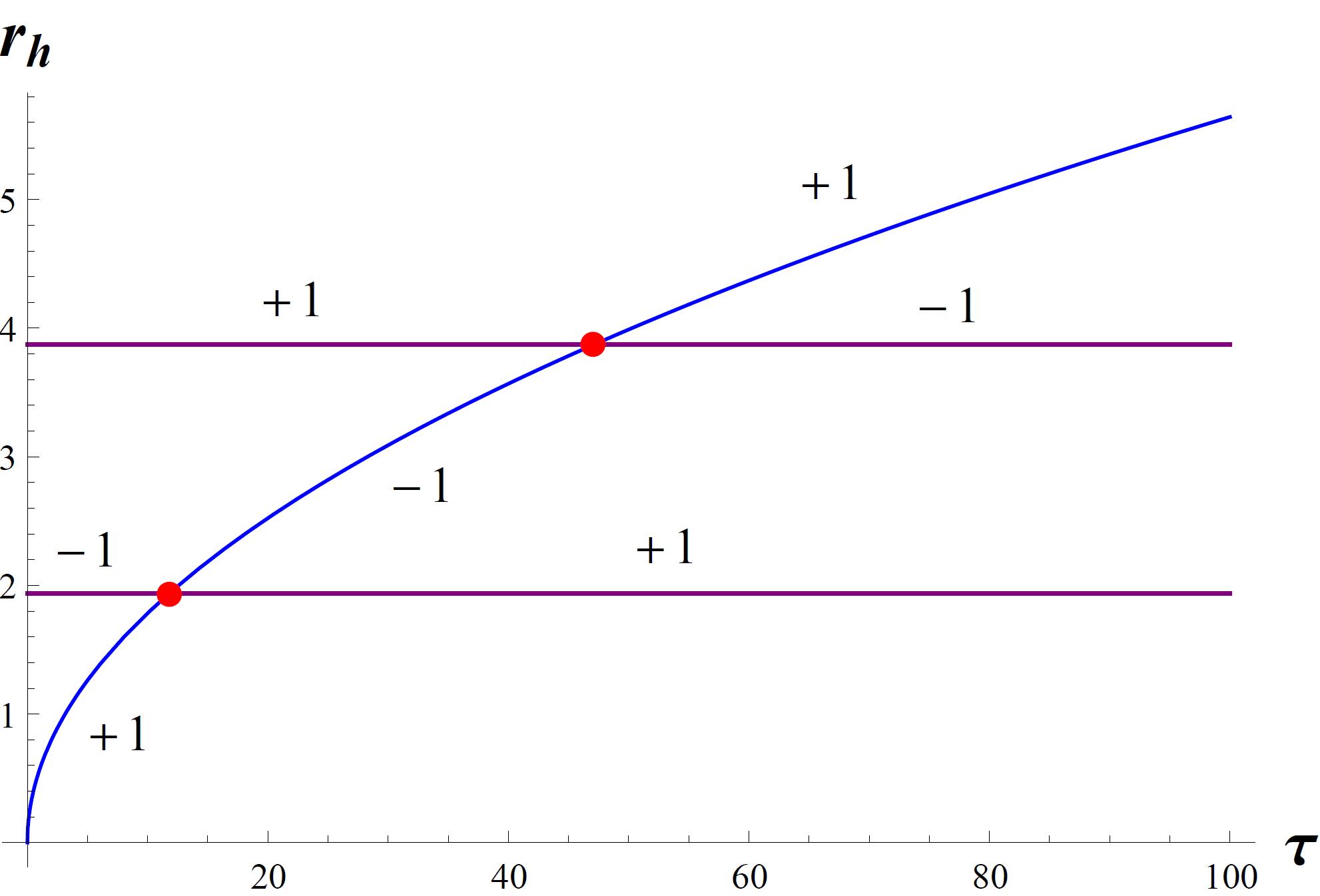}
  \caption{Defects for charged black holes in AdS. The blue/horizontal line stands for black holes/exotic defects. $\pm 1$ stands for the local winding nubmer associated to each solution. The red points denote the bifurcation points $r_h^*=r_{h\pm}$ and $\tau^*=\pi r_{h}^{*2}$, at which black holes collide with the exotic defects. Critical point of the small-large black hole transition can be extracted when all the defects meet in the space.}\label{rn}
\end{figure}

It turns out that critical point of the transition can be extracted when the two exotic defects merge into one, i.e. $r_{h+}=r_{h-}$ or $\partial_{r_h}J^1=0$ directly. We find $ r_{h\,,c}=\sqrt{6}\,Q\,,T_c=1/3\sqrt{6}\pi Q\,,P_c=1/96\pi Q^2$.
The result is exactly the same as that obtained from the infection point of isotherms in $P-r_h$ diagram \cite{Kubiznak:2012wp}. As a cross-check, we directly calculate the velocity at the critical bifurcation point. We find Eq.(\ref{bif}) becomes invalid owing to a higher degeneracy. By expanding $\phi^1(\tau\,,r_h)$ to cubic order, we deduce
\be A_1 \Big( \fft{dr_h}{d\tau} \Big)^3+3A_2 \Big( \fft{dr_h}{d\tau} \Big)^2=0\,,\ee
where $A_1=1/\sqrt{6}Q^3\,,A_2=-1/36\pi Q^4$. It turns out that there are three branch solutions, two of which have a vanishing velocity, corresponding to the exotic defects. The third solution has
$u^1=-3A_2/A_1=1/2\pi r_{h\,,c}$, exactly corresponding to the critical black hole.

Near the critical point, the Jacobians behave as $J^1\propto t/18\pi Q^2\,,J^0\propto t/9Q$, where $t=\tilde T-1$. This captures the critical behavior of isothermal compressibility \cite{Kubiznak:2012wp}. Besides, from velocity, we deduce $ u_s-u_\ell=2u_c\sqrt{-2t}$. This characterizes the critical behavior of specific volume of black holes near the critical point.

{\it Conclusions.--}By employing an off-shell internal energy, we have developed a universal picture from topology in the thermodynamic parameters space to describe first order transitions of black holes. In our prescription, there are two types of defects. The first is normal, describing black holes, which move with a nonzero velocity and acceleration in a central force field. The second type is static and exotic, not describing black holes but encoding information about first order transitions.

By studying bifurcation points in the parameters space, we illustrate that first order transitions can be interpreted as once or twice interchange of winding numbers between the defects through wired action at a distance. However, the process can also be locally interpreted as virtual collisions between the normal defects and the exotic ones. In this interpretation, the defects simply change their positions and momentums rather than interchanging the winding numbers. Critical point of the transition can be extracted when all the defects meet in the parameters space.


 Our discussions for black holes are readily extended to general thermodynamic systems by identifying $F$ to the Gibbs free energy, $r_h$ to the specific volume and all derivatives are understood to be evaluated at constant pressure. Without presenting any details, we claim that the same picture hold for the liquid-gas transition of Van der Waals fluids. Generalization to first order transitions between multi-phases will be very interesting.

It is also possible to generalize our approach to second order transitions. In this case, the exotic defects are irrelevant. Recall the topological charge density $j^0=\delta^2(\vec\phi)J^0$, where $J^0=4\pi TS/C_T$. According to the mean field theory, the specific heat $C_T$ jumps discontinuously at the critical point of second order transitions. This implies that the topological charge density $j^0$ will jump discontinuously at the critical point as well. Since the two phases are generally described by different topological charge density, we may interpret the breaking or restoration of a continuous symmetry during second order transitions as the sudden changes of local distribution of the topological charge density $j^0$. Note the topological charge is unaffected during the transition, significantly different from first order transitions. As an example, for magnetic monopoles, it means a second order transition occurs between two phases which have the same magnetic flux but different magnetic fields.

{\it Acknowledgments.--} This work is supported in part by the National Natural Science Foundations of China (NNSFC) with Grant No. 11873025. \\

\end{document}